\documentclass[
 preprint,
 amsmath,amssymb,
 aps,
 prd
]{revtex4-1}

\begin{document}

\title{The effect of Earth's gravitational field on the muon magic momentum}

\author{Pawel Guzowski}
\affiliation{
 The University of Manchester, School of Physics and Astronomy, Manchester M13 9PL, United Kingdom
}

\date{\today}

\begin{abstract}

A recent set of articles considers the effect of Earth's gravity on the magnetic moments of fermions. The authors conclude that the gravitational effects cancel out for measurements of the electron anomalous magnetic moment, but for the muon measurements, given that they are performed on highly relativistic muons, the effect is large enough to explain the discrepancy between measured and theoretical values. In this paper this effect is shown to be too small given experimental resolutions, and has been absorbed into the systematic uncertainties of experiments performed to date. Future experiments will require muon momentum tuning at the keV/$c$ level to see this effect.

\end{abstract}

\maketitle

The authors of Ref~\cite{mfs} claim that the discrepancy between the muon anomalous magnetic moment measured by the BNL experiment~\cite{bnl} and the theoretical value~\cite{rpp} can be explained by the general relativistic curvature of spacetime on the surface of the Earth. They say that \emph{if} the anomalous magnetic moment was measured by tuning the ``magic'' $\gamma$ of the experiment to minimize the electric field contribution, then the measured value would differ from the true value by $2.8\times 10^{-9}$, which is the same size as the experimental discrepancy.

However, in prior and current experiments, the anomalous magnetic moment \emph{is not} measured by tuning the magic momentum, but by using the ratio of various measured precession frequencies, where the gravitational effects cancel out (as explained in section 2.5 of Ref~\cite{mfs}).

In addition, the change in the magic $\gamma$ due to the gravitational effect of the Earth is too small to be significant in current experiments.

In a flat spacetime, the anomalous precession frequency $\omega_a$ is
\begin{equation}
\vec{\omega}_{a} = -\frac{q}{m}\left[a_\mu \vec{B}-\left(a_\mu-\frac{1}{\gamma^2-1}\right)\frac{\vec{\beta}\times\vec{E}}{c}\right],
\end{equation} where $a_\mu$ is the anomalous magnetic moment, $q$ the elementary electric charge, $m$ the muon mass, $\vec{B}$ the magnetic field, $\vec{E}$ an electric field, $\vec{\beta}$ the muon velocity, $\gamma$ the Lorentz factor of the muon, and $c$ the speed of light~\cite{mfs,bnl}. The ``magic'' Lorentz factor for a flat spacetime, $\gamma_0$, is chosen such that the electric field contribution to $\omega_a$ cancels,
\begin{equation}
\gamma_0^2 = \frac{1+a_\mu}{a_\mu}.
\end{equation}

The calculations in Ref~\cite{mfs} for the curved spacetime give the modified anomalous precession frequency, assuming that the magnetic field is perpendicular to the muon motion, as
\begin{equation}
\vec{\omega}_{a} = -\frac{q}{m}\left[\left(1+3\epsilon^2\phi\right)a_\mu \vec{B}-\left(a_\mu-\frac{1}{\gamma^2-1}-\epsilon^2\phi\left(4+a_\mu+\frac{3}{\gamma^2-1}\right)\right)\frac{\vec{\beta}\times\vec{E}}{c}\right],
\end{equation}
where $\epsilon=1/c$, $\phi=-G M/r$, $G$ the Newtonian gravitational constant, $M$ the mass of the Earth, and $r$ the radius of the Earth. In this case the magic Lorentz factor for a curved spacetime, $\gamma_\phi$, is calculated as
\begin{equation}
\gamma^2_{\phi} = \frac{\left(1+a_\mu\right)\left(1-\epsilon^2\phi\right)}{a_\mu\left(1-\epsilon^2\phi\right)-4\epsilon^2\phi}.
\end{equation}

For a muon with $a_\mu=1.16591803\times10^{-3}$, $\gamma_0=29.3$ and the magic momentum $p_0= m \sqrt{\gamma_0^2-1}=3.1~\textrm{GeV}/c$.
For the Earth, $\left|\epsilon^2\phi\right| = 6.9\times10^{-10}$, and $\left|\gamma_{\phi}-\gamma_0\right|=1.2\times10^{-6}\gamma_0$. The magic momentum is modified by only $\sim 3~\textrm{keV}/c$. The magic radius (the radius of the muons at the magic momentum in the BNL experimental setup) is changed by only around 10 microns, smaller than the 500 micron uncertainty of the placement of the electric quadrupoles in the BNL experiment~\cite{bnl}.

The range of momenta in the storage ring of the BNL experiment is represented in Figure~20 of Ref~\cite{bnl}, showing a momentum spread around the magic value of around $0.15\%$, much greater than the $\sim 1~$ppm due to gravitational effects. 
Muons of the modified magic momentum would have been present in the experiment, and the analysis already takes into account~\cite{bnl} that some muons will be travelling slightly faster or slower than the magic momentum, with an associated systematic uncertainty on the final measurement. 

Corrections are applied to the BNL $\omega_a$ measurement in order to correct for muons not travelling horizontally  (the pitch correction) or at the magic momentum (electric-field correction). These will be over or under corrected if the full gravitational effects are taken into account. The pitch correction should be itself corrected by a factor of $(1-\epsilon^2\phi(2\gamma-1))$ according to the calculations of Ref~\cite{mfs}, which is approximately $(1+10^{-8})$ and so is negligible. The correction factor needed for the electric-field correction is $(1-\epsilon^2\phi-4\epsilon^2\phi/a_\mu)\simeq(1+10^{-9}+10^{-6})$, and so again is negligible. This is a 1 ppm multiplicative correction factor to the correction itself, not a 1 ppm additive absolute value to the overall correction.

In conclusion, the effect of Earth's gravity on the magic momentum of the muon is much smaller than the experimental sensitivity for past and current experiments. The g-2 experiment~\cite{g-2} has a similar design to BNL, with a similar momentum acceptance, and so is unlikely to see this effect. Future experiments would need a muon momentum tuning resolution at the keV/$c$ level for this effect to be seen, and the beam would have to be confined to a much tighter radius and with more precise determination of experimental apparatus alignments.

\end{document}